\documentclass[twocolumn,showpacs,preprintnumbers,amsmath,amssymb]{revtex4}
\usepackage{graphicx}
\usepackage{dcolumn}
\usepackage{bm}

\begin{document}

\preprint{}

\title{A new method of observing\\ weak extended x-ray sources with RHESSI}

\author{Iain G. Hannah}
\email{hannah@ssl.berkeley.edu}
 \author{Gordon J. Hurford}
  \author{Hugh S. Hudson}
   \author{Robert P. Lin}
\affiliation{%
Space Sciences Laboratory, University of California at Berkeley, \\Berkeley, CA,
94720-7450, USA
}%

\date{\today}

\begin{abstract}
We present a new method, fan-beam modulation, for observing weak extended x-ray
sources with the Reuven Ramaty High-Energy Solar Spectroscopic Imager (RHESSI).
This space-based solar x-ray and $\gamma$-ray telescope has much greater
sensitivity than previous experiments in the 3-25 keV range, but is normally not well
suited to detecting extended sources since their signal is not modulated by RHESSI's
rotating grids. When the spacecraft is offpointed from the target source, however,
the fan-beam modulation time-modulates the transmission by shadowing resulting
from exploiting the finite thickness of the grids. In this paper we detail how the
technique is implemented and verify its consistency with sources with clear known
signals that have occurred during RHESSI offpointing: microflares and the Crab
Nebula. In both cases the results are consistent with previous and complementary
measurements. Preliminary work indicates that this new technique allows RHESSI to
observe the integrated hard x-ray spectrum of weak extended sources on the quiet
Sun.
\end{abstract}

\pacs{95.55.Ev,95.55.Ka,95.75.-z}

\maketitle

\section{Introduction}

The Reuven Ramaty High-Energy Solar Spectroscopic Imager, RHESSI \citep{lin2002},
is a space-based solar x-ray and $\gamma$-ray telescope that was launched in
2002. Its main goal is to obtain spectroscopic and imaging information of solar flares
with high time and energy resolutions from 3~keV to 15~MeV. In particular it has
unprecedented sensitivity for 3-25~keV x-rays.  This is because when its automated
attenuators (used in flare observations) are ``out'', it can observe down to the
nominal 3~keV limit with the full area of the detectors, something not possible for
earlier instruments which used fixed shielding to prevent excessive counting rates
from soft x-rays in flares. Normal RHESSI imaging is accomplished with a set of nine
bigrid rotating modulation collimators (RMCs) with resolutions logarithmically
spaced from 2.3$''$ to 183$''$. Each RMC time-modulates sources whose size scale
is smaller than their resolution. Thus despite its sensitivity, it is not well suited to
observe weak  sources larger than 3 arcminutes. For weak sources it is also
essential to distinguish counts due to photons from the target from those due to
terrestrial, cosmic or instrumental background.

To achieve this we have developed a technique called \emph{fan-beam modulation},
detailed and tested in this paper, that involves pointing the telescope slightly away
from the target. As RHESSI rotates, the narrow field of view ($\sim 1^\circ$ FWHM)
of RHESSI's thick grids time-modulates the signal to ``chop''  between an extended
source and background.

The main motivation for developing this technique is to obtain the hard x-ray
spectrum of the Sun free of sunspots, active regions and flares (the quiet Sun). We
expect the quiet Sun sources to be weak and well-dispersed across the solar disk
(the diameter of which is about 32$'$) . Since hard x-ray instrumentation is typically
optimized for flare observations (bright compact sources), such quiet Sun
observations remain an elusive measurement despite interest back to the earliest
days of solar x-ray observations \citep[e.g.,][]{neupert1969}. With RHESSI we have
the possibility of improving these values as well as extending them to lower and
higher energies.

The focus of this paper is to introduce the \emph{fan-beam modulation} technique,
whose implementation is explained in Sec. \ref{sec:offpnt}. In Sec. \ref{sec:test} we
test this method on sources with strong or well-known signals: microflares that
occurred during quiet Sun offpointing, and the Crab Nebula. In Sec. \ref{sec:qs} we
present some analysis from periods of low solar activity, illustrating that this method
is capable of observing weak extended sources.

\begin{figure*}\centering
\includegraphics[]{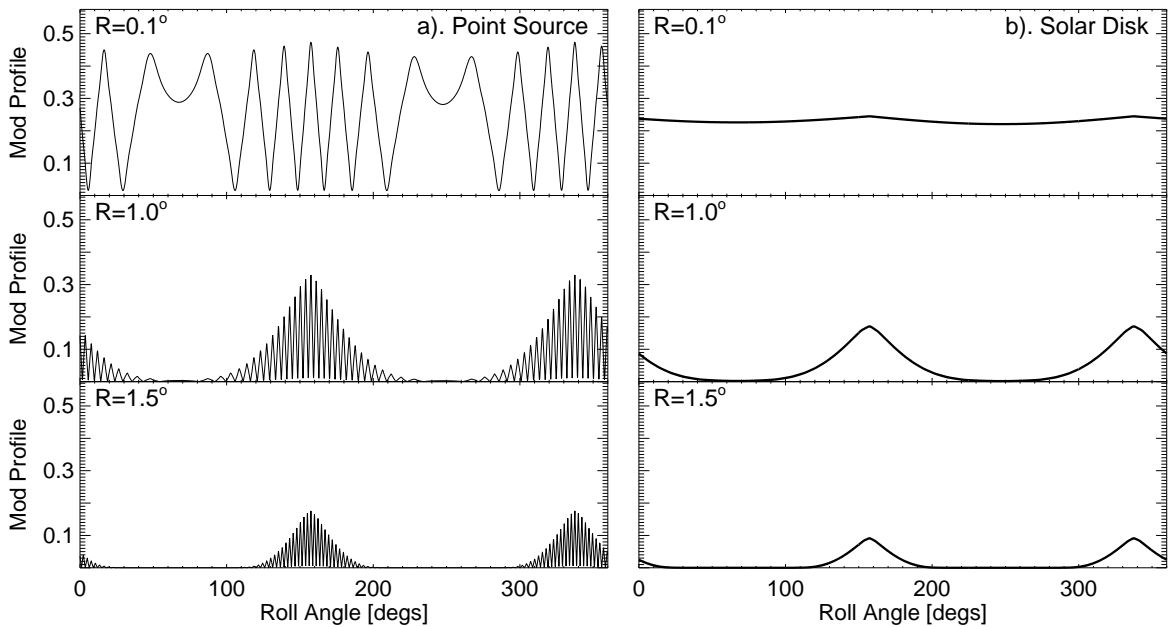}
\caption{ The expected modulation profile of a a). point source and the b). solar disk
through one of RHESSI's rotating collimators, for three different radial offsets
between RHESSI pointing and source center. Normal RHESSI imaging is based on the
rapid modulation in the left panel.}\label{fig:modprof}
\end{figure*}

\section{\label{sec:offpnt}RHESSI Offpointing}
The RHESSI telescope consists of nine pairs of grids in front of nine germanium
detectors \citep{lin2002}, which record x-rays with high energy and time resolution
from 3~keV to 15~MeV. To obtain temporal modulation of the counting rates the
spacecraft rotates about the axis pointing at the Sun with a rotation period of about
4~seconds. Each pair of grids then constitutes a rotating modulation collimator, RMC
\citep{1968SSRv....8..534Schnopper}. Compact sources on the solar disk (such as
flares) are rapidly time-modulated by the grids, enabling high-resolution imaging,
but the primary modulation of solar-size sources is negligible.

There is, however, a secondary modulation that results from the finite thickness of
the collimator grids \citep{hurford2002}. This ``envelope'' modulation peaks twice
every rotation when the slits of the grids are parallel to the line between RHESSI
pointing and source center, producing two transmission maxima per rotation. For one
collimator, Fig. \ref{fig:modprof} shows the expected modulation profiles for a point
source and for the solar disk as a function of roll angle for several off-axis angles.
The top row of this figures shows the result for typical solar pointing: a point source
has the expected rapid modulation, but this does not occur for the full solar disk. As
the offpointing angle increases the frequency of the rapid modulation increases as
well as the secondary fan-beam modulates the signal twice per rotation.

\begin{figure}\centering
\includegraphics[]{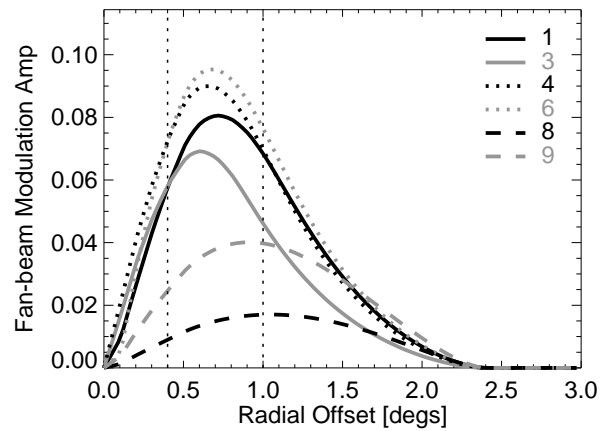}
\caption{The fitted amplitude of the fan-beam modulation (for a uniform solar disk)
as a function of radial offset from the target. The modulation is maximised at offsets
from 0.4$^\circ$ to 1.0$^\circ$.} \label{fig:gridtranrd}
\end{figure}

\begin{figure}\centering
\includegraphics[]{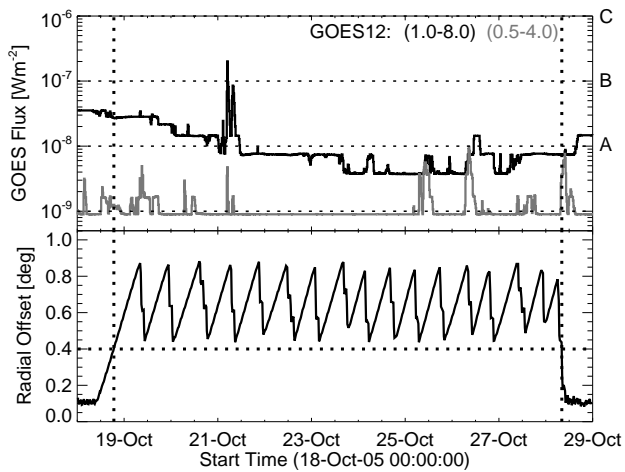}
\caption{Summary of RHESSI offpointing during October 2005. The top panel shows
the GOES-12 x-ray flux in the 1-8\AA~and 0.5-4\AA~channels, the bottom panel the
radial offset of RHESSI pointing from Sun center.}\label{fig:octsumry}
\end{figure}

The fan-beam modulation increases with the offpointing angle as shown in
Fig.~\ref{fig:gridtranrd}, where we have plotted its amplitude as a function of radial
offset. For this work we do not use the data from RMCs 2, 5 or 7 since their
thresholds are above 3~keV and/or have degraded energy resolution. The first four
RMCs shown have the smallest field of view ($\sim1^\circ$) orthogonal to the slits
and so the modulation is greater than in RMCs 8 and 9, whose FOV is $\sim8^\circ$
and $\sim3^\circ$ respectively. Fig.~\ref{fig:gridtranrd} also shows that we get the
maximum effect of the fan-beam modulation technique when RHESSI is pointing
between $0.4^\circ$ and $1^\circ$ away from source center.  At larger angles the
increasing shadowing of the grids on the detectors reduces the amplitudes. Normally
the RHESSI pointing control system remains inactive when its rotation axis is within
a dead-band corresponding to solar offsets between 0.05$^\circ$ and 0.2$^\circ$.
To initiate offpointing the dead-band limits are changed to 0.4$^\circ$ and
1.0$^\circ$. For quiet Sun observations, this mode of operation has been initiated
several times when the GOES soft x-ray flux was low and no active regions or spots
were on the disk. A summary of RHESSI's pointing and the GOES flux during the
October~2005 offpointing is shown in Fig. \ref{fig:octsumry}.

To facilitate the analysis of such a large data set it is divided into successive
five-minute time intervals, which are short enough that the pointing of the
spacecraft changes little. The data for each five-minute interval can be analyzed in
any energy range above~3~keV, although smaller energy bands provide poorer
statistics. For a chosen energy range we bin (``stack'') the data according to the roll
angle, $\theta$, for each subcollimator, correcting for the relative geometry of the
spacecraft pointing and the source and the grid orientation such that $\theta=0$
when the grid slits are aligned to the source. We then fit this stacked data with
\begin{equation}\label{eq1}
F(\theta)=A_\mathrm{0}+A_\mathrm{1}\cos(2\theta+\phi_\mathrm{1})+A_\mathrm{2}\cos(4\theta+\phi_\mathrm{2}),
\end{equation}
where A$_\mathrm{0}$ is the background term, A$_\mathrm{1}$ and
A$_\mathrm{2}$ are the main and first harmonic amplitudes of the signal and
$\phi_{n}$ are the fitted phases. The source signal should occur at $0^\circ$ and
$180^\circ$, with the phase offset $\phi$ should be zero, as we have corrected
$\theta$. The A$_\mathrm{1}$ amplitude then provides the measure of the source
flux although it has to be corrected for the grid transmission efficiency and grid
shadowing, factors dependent on radial offset shown in Fig. \ref{fig:gridtranrd}. After
this correction, the resulting fits from different time intervals can then be combined
vectorially to improve the signal-to-noise ratio before conversion to a final photon
flux. This conversion is achieved by either using the diagonal elements of the
appropriate detector response matrix \citep{smith2002}, or via forward-fitting a
model to the spectrum using the full detector response matrix in the OSPEX software
package (an updated version of the SPEX code written by R. Schwartz).

\begin{figure}\centering
\includegraphics[]{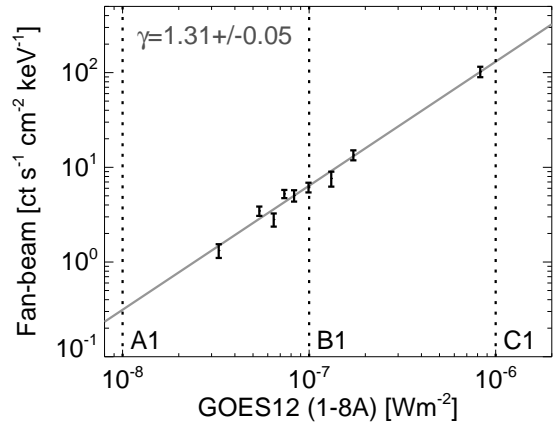}
\caption{The mean counts flux in 4-6 keV obtained using the fan-beam modulation
technique for nine microflares occurring during quiet Sun offpointing, compared to
the background subtracted GOES 1-8\AA~flux. }\label{fig:flvsgs}
\end{figure}

\begin{figure}\centering
\includegraphics[]{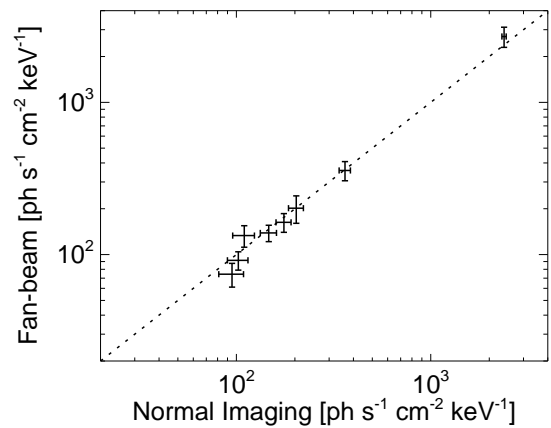}
\caption{Comparison of the mean 4-6~keV photon flux, obtained using the fan-beam
modulation technique, and the photon flux found through a normal RMC9
back-projection image, for 8 microflares (imaging was not possible for one of the 9
microflares shown in Fig.~\ref{fig:flvsgs}).}\label{fig:flvsnorm}
\end{figure}

\begin{figure*}\centering
\includegraphics[]{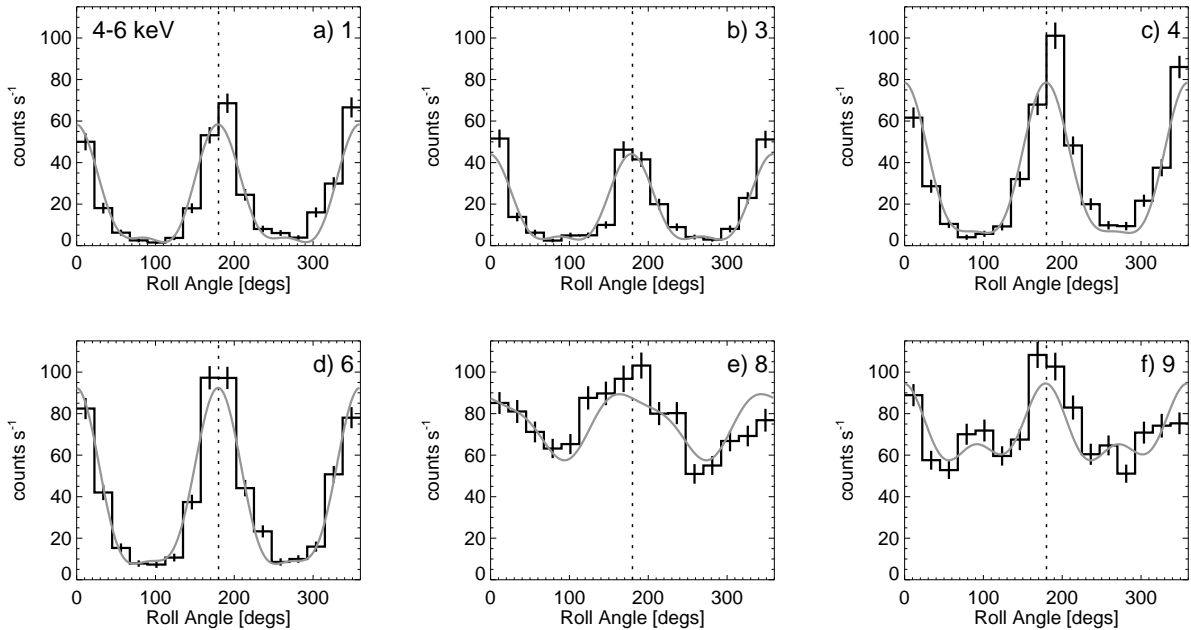}
\caption{The stacked RHESSI counts in 4-6 keV through each suitable subcollimator
for an A8.3 microflare. In this fit we have included the first harmonic as well as the
fundamental term. Since the roll angle has been corrected for the orientation of the
grids and relative positions of RHESSI pointing and the microflare, the signal should
peak at $0^\circ$ and $180^\circ$.}\label{fig:stackfl}
\end{figure*}

\section{Testing the Offpointing Procedure}\label{sec:test}

We test the fan-beam modulation technique on microflares which have a relatively
large signals, and the Crab Nebula, an extended source whose weaker signal is well
known.

\subsection{Microflares}\label{sect:offp}

Several microflares occurred during quiet sun offpointing intervals. One typical event
occurring at 22:33:00 25 July 2005 was equivalent to a GOES A8.3 flare with
background subtracted. One minute of 4-6~keV stacked data for this event is shown
in Fig. \ref{fig:stackfl}. In RMCs 1, 3, 4 and 6 we get a strong and clearly defined
signal, which matches the expected modulation. In RMCs 8 and 9 the modulation is
not as strong, as expected from Fig.~\ref{fig:gridtranrd}. Making the grid
transmission correction to the amplitude fit of the signal in RMCs 1, 3, 4 and~6 we
average the four fluxes to obtain a single value for this flare in the 4-6~keV band.

This analysis was repeated for each of the nine microflares for which we achieved
good coverage during our four quiet-Sun offpointing periods. Fig.~\ref{fig:flvsgs}
shows the RHESSI 4-6~keV microflare counts, corrected for grid transmission,
against the corresponding GOES 1--8\AA~energy fluxes. The errors shown for the
values derived from the fan-beam modulation are based on the standard deviation of
the values found from each of the four individual detectors. This systematic error
among the detectors (about 13.5\% for these microflares) is larger than the
statistical error found from each of the fits. This systematic discrepancy is a known
issue with RHESSI and is still under investigation. The correlation between the
RHESSI and GOES fluxes shown in Fig.~\ref{fig:flvsgs} is an excellent fit to a power
law but differs from unity (power-law index~1.33). This may reflect temperature
differences among flares convolved with the difference between the GOES and
RHESSI spectral response.

When RHESSI is offpointed from the Sun, pitch and yaw information is obtained from
the Fine Sun Sensor (FSS) which is intrinsically less precise than the Solar Aspect
System (SAS) used for conventional RHESSI imaging. However its accuracy is more
than sufficient to support fan-beam modulation analyses as well as conventional
back projection imaging with the coarsest subcollimator, RMC9. The agreement
between conventional imaging and the fan-beam modulation technique is illustrated
in Fig.~\ref{fig:flvsnorm} which shows a comparison of the 4-6~keV fluxes.

\subsection{Crab Nebula}\label{sect:crab}

\begin{figure}\centering
\includegraphics[]{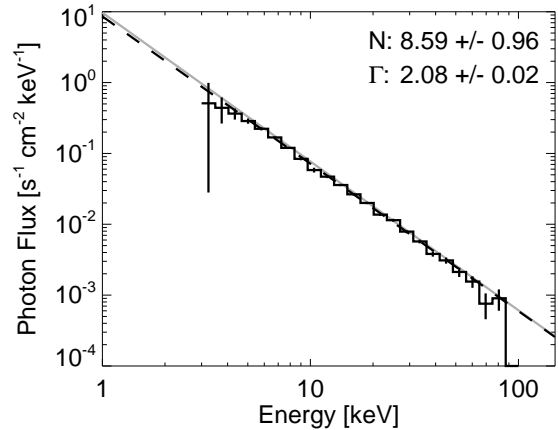}
\caption{The spectrum of the Crab found using the fan-beam modulation technique
(histogram). The dashed black line is a powerlaw fit to the data between 4~and
60~keV. The grey line is the reference spectrum from Toor and
Seward\cite{toor1974}.} \label{fig:crabspec}
\end{figure}

In an annual program, RHESSI points away from the Sun to image the Crab Nebula
and pulsar, around the time of the Sun's closest approach (mid-June). Since the hard
x-ray spectrum of the Crab is well-known and stable \citep{kirsch2005} it provides a
photometric test of the fan-beam modulation offpointing technique.

\begin{figure*}\centering
\includegraphics[]{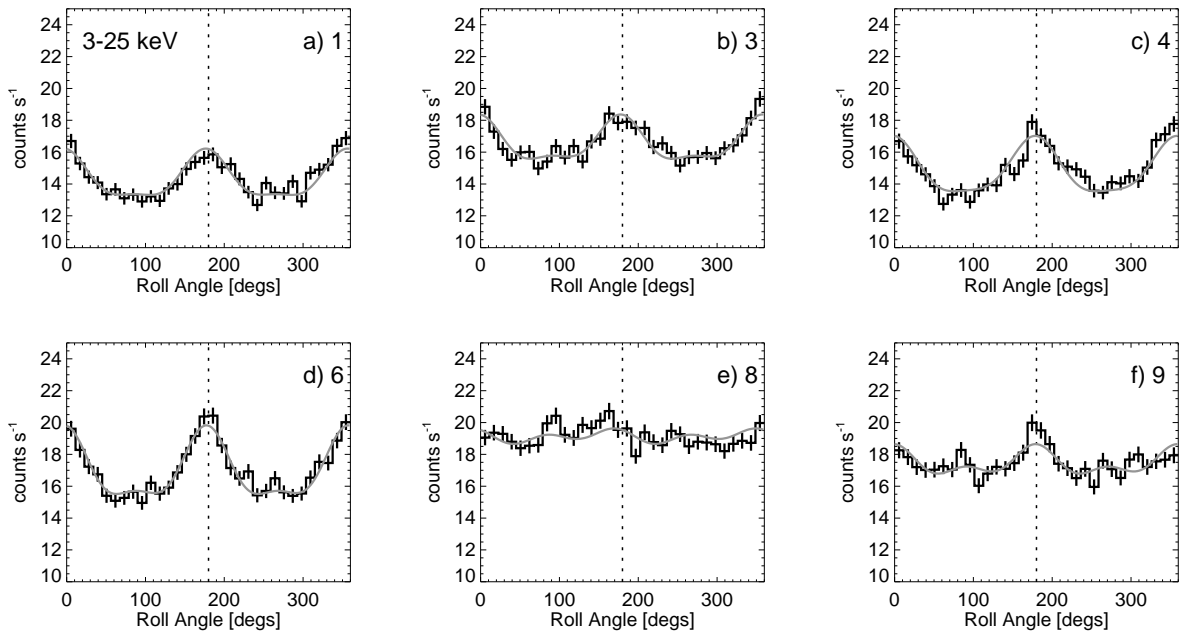}
\caption{The stacked 3-25 keV RHESSI counts for an hour-long integration during
Crab offpointing in June 2004. The roll angle has been corrected for the orientation
of the grids and relative positions of the Crab and RHESSI pointing, so that the Crab
signal should be at $0^\circ$ and $180^\circ$. The continuous line shows the model
fit for a point source at the location of the Crab Nebula. This data have not been
corrected for grid dependent parameters such as thickness.}\label{fig:stackcrb}
\end{figure*}

In the 2004 campaign, some data were taken at radial offsets suitable for testing the
fan-beam modulation technique (the Crab between $0.4^\circ$ and $1^\circ$ away,
the Sun $>2.5^\circ$). As a result the Crab signal is modulated while the Sun is too
far away to be directly observed. Thus even in the presence of solar activity we can
obtain a clean Crab signal. From 11 to 20 June 2004 this orientation occurred for a
total of 19 such orbits (each about an hour in length), avoiding times of high particle
background during the South Atlantic Anomaly (SAA) passages.

The stacked counting rates in the 3-25~keV band for one of these orbits is shown in
Fig.~\ref{fig:stackcrb}. As before the roll angles are corrected for the grid orientation
and the relative position of the Crab to RHESSI pointing, so that the Crab signal
should occur at $0^\circ$ and $180^\circ$. Fig.~\ref{fig:stackcrb} shows the stacked
signal for six of the RHESSI detectors. In RMCs 1, 3, 4 and 6 we get the expected
modulation which is fitted well by Eq. \ref{eq1} (shown as a grey line).

For each of the four detectors we fit the stacked count-rate data for each orbit in a
series of energy bands from 3~to 100~keV. The amplitude of these fits is corrected
for the grid transmission for the particular radial offset of that orbit. This provides a
count spectrum, corrected for grid transmission, for each of the four detectors. Then
each of these spectra is forward-fitted using the OSPEX software package
\citep{schwartz2002} with a power law and the appropriate detector response
matrix, to obtain a photon spectra. The four photon spectra and their fits are then
combined and shown in Fig.~\ref{fig:crabspec}.

Fitting a power law $dN/dE = N E^{-\Gamma}$ ph~s$^{-1}$ cm$^{-2}$ keV$^{-1}$,
we obtain $N=8.59~\pm~0.96$ and $\Gamma=2.08~\pm~0.02$, for the RHESSI data
between 4 and 60 keV, as shown in Fig.~\ref{fig:crabspec}. The standard values for
these parameters are $N=9.7 \pm 1.1$, $\Gamma=2.1 \pm 0.03$ for the range
2--50~keV \citep{toor1974}. The consistency with our observations provides
confidence in their photometric accuracy. Comparison with a more recent survey of
these parameters also provides similar results \citep{kirsch2005}.

\begin{figure*}\centering
\includegraphics[]{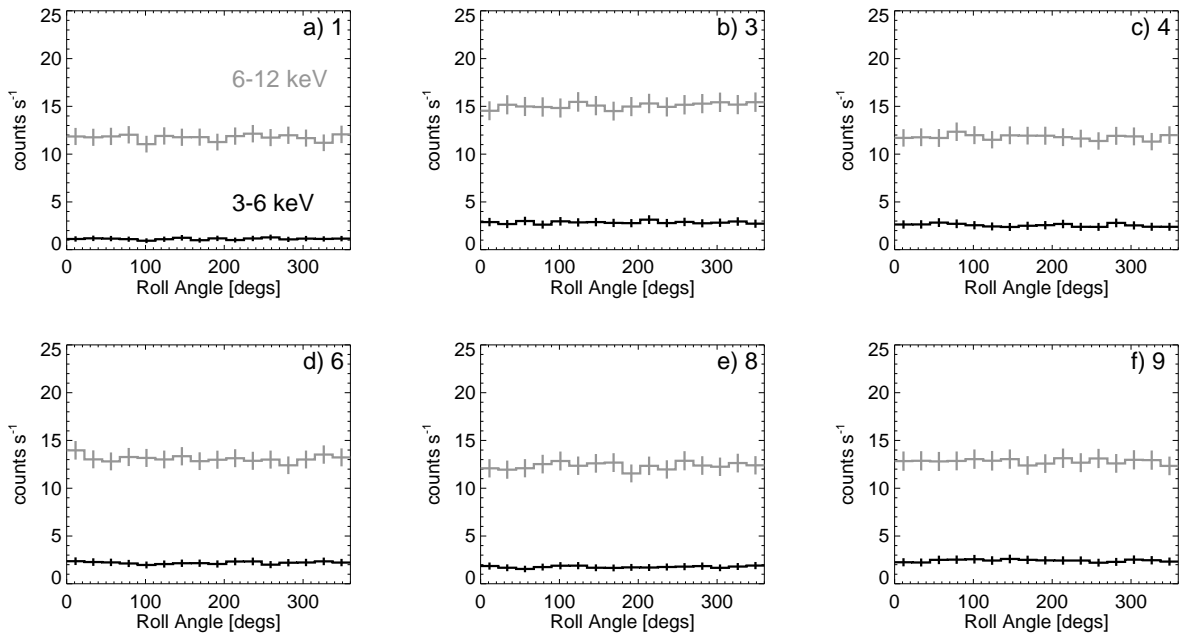}
\caption{The stacked RHESSI counts through each suitable subcollimator for 1 hour
of quiet Sun during offpointing. Two energy bands are shown: 3-6 keV (black) and
6-12 keV (grey). Note that the roll angle has been corrected for the orientation of the
grids and relative positions of the Sun and RHESSI pointing, so that the solar signal
should be at $0^\circ$ and $180^\circ$.}\label{fig:stackqs}
\end{figure*}

\section{\label{sec:qs}Test on the non-flaring Sun}

Here we present a preliminary analysis of the quiet Sun signal, to indicate that this
method works on the intended target source.

For this analysis we have used the data from the four initial quiet Sun offpointing
periods, as detailed in Sec.~\ref{sec:offpnt} and illustrated in Fig.~\ref{fig:octsumry}.
During this period the Sun was very quiet with the GOES 12 flux in 1-8\AA
``flat-lining'' below $10^{-8}$~Wm$^{-2}$. We do not use the full offpointing periods
in the analysis; instead we screen the time intervals by checking the RHESSI data for
sharp time-series features (such as flares or particle events). We have also
restricted the data to times when RHESSI is at the lowest latitudes in its orbits, to
minimize the terrestrial background. From these four offpointing periods we have a
total of 1,522 five-minute time intervals (over 126 hours of data).

For each of the five-minute intervals we selected a set of energy bins and fit the
stacked data in each subcollimator with Eq. (\ref{eq1}). The stacked data in two
energy ranges (3-6~keV and 6-12~keV), for 1 hour of quiet Sun data, with GOES
background flux below A1~level, are shown in Fig.~\ref{fig:stackqs}.  No obvious
signal appears, but none would be expected. This highlights the need to combine
many time intervals to improve the signal-to-noise. We fitted the stacked data from
each RMC with Eq.~\ref{eq1} in such a way that an $A_{i}\cos\phi_{i}$ and
$A_{i}\sin\phi_{i}$ term, with associated errors, is returned for each of the cosine
terms in Eq.~\ref{eq1}. Statistically the expectation value of
$\left<A_\mathrm{1}\cos\phi_\mathrm{1}\right>\approx A_\mathrm{1}$, the source
signal, and $\left<A_\mathrm{1}\sin\phi_\mathrm{1}\right>\approx 0$. The former
represents the measurement of interest while the latter provides a consistency
check.

\begin{figure}\centering
\includegraphics[]{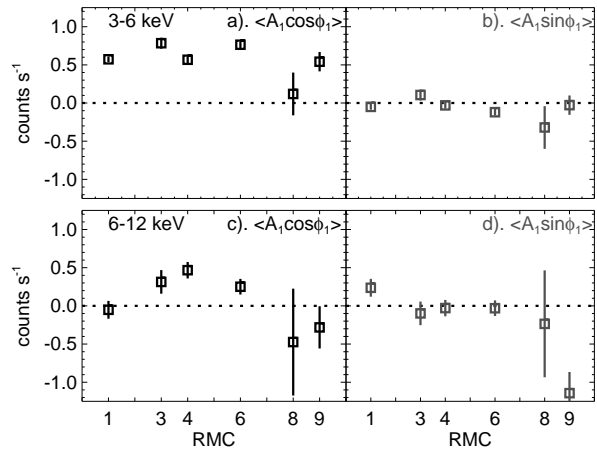}
\caption{The mean value of the fit components
$A_\mathrm{1}\cos\phi_\mathrm{1}$ (left panels, a). and c).) and
$A_\mathrm{1}\sin\phi_\mathrm{1}$ (right panels, b). and d).) for detectors RMCs 1,
3, 4, 6, 8 and 9, averaged over the 1,522 five-minute quiet Sun time intervals. The
top row show the results for 3-6 keV, the bottom row for 6-12
keV.}\label{fig:qsmeans}
\end{figure}

Fig.~\ref{fig:qsmeans}a shows the mean 3-6~keV counts (corrected for grid
transmission) for each detector averaged over all 1,522 five-minute intervals. The
larger error bars for RMC8 and 9 reflect their lower modulation efficiency. Although
the scatter is larger than the statistics would suggest, there is a clear detection.
Fig.~\ref{fig:qsmeans}b shows the corresponding average for the
$\left<A_\mathrm{1}\sin\phi_\mathrm{1}\right>$ term which, as expected, is
consistent with zero. In the higher 6-12~keV energy channel we achieve similar
results, with the $\left<A_\mathrm{1}\cos\phi_\mathrm{1}\right>$ positive and
non-zero and $\left<A_\mathrm{1}\sin\phi_\mathrm{1}\right>$ consistent with zero.
These results both show that the fan-beam modulation technique works by being
able to detect a signal from the non-flaring quiet Sun.

\section{\label{sec:cons}Conclusions}

We have developed a new technique for observing the quiet Sun which uses of the
fan-beam collimation of the grids. The technique extends RHESSI observations to
sources larger than can be imaged using conventional bigrid modulation. As
presented in Sec.~\ref{sect:crab}, the photometric accuracy of the technique has
been verified using the known flux of the Crab Nebula.

From the initial tests of the fan-beam modulation technique on the quiet Sun we are
able to achieve positive non-zero signal. We would then expect that the fan-beam
modulation technique holds promise for offpointing periods when observations in
the magnetically quieter conditions expected in 2007.
\begin{acknowledgments}
NASA supported this work under grant NAG5-12878. We thank many people involved
in the RHESSI program for inspiration and assistance with software, calibration, and
spectral fitting issues, especially Brian Dennis, Richard Schwartz and David Smith.
\end{acknowledgments}


\end{document}